\documentclass[fleqn]{tlp}

\newtheorem{example}{Example}[section]
\newtheorem{definition}{Definition}[section]
\newtheorem{theorem}{Theorem}[section]
\newtheorem{algorithm}{Algorithm}[section]

\newcommand{\sssection}[1]{\vspace{-3mm}\subsubsection*{#1}\vspace{-2mm}}
\newcommand{\LDL}{\mbox{$\cal LDL$}}
\newcommand{\tz}{\tt theta}
\newcommand{\bldl}{\smallskip\[\begin{array}{ll}\small}
\newcommand{\cldl}{\vspace{-0.5cm}\[\begin{array}{ll}\small}
\newcommand{\eldl}{\end{array}\]\rm}
\newcommand{\prule}[2]{\tt #1 \leftarrow & \tt #2 \\}
\newcommand{\pfact}[2]{\tt #1 & \tt #2 \\}
\newcommand{\pbody}[2]{\tt #1 & \tt #2 \\}
\def\<{\mbox{$\langle$}}
\def\>{\mbox{$\rangle$}}

\title[Greedy Algorithms in Datalog]{Greedy Algorithms in Datalog}

\author[S. Greco and C. Zaniolo]
{Sergio Greco \\
Dip.  Elettronica  Informatica e Sistemistica \\
Universit\`a  della  Calabria \\
87030 Rende, Italy  \\
greco@si.deis.unical.it \\
\and Carlo  Zaniolo \\
Computer Science  Department \\
University of California at Los Angeles\\
Los Angeles, CA 90024 \\
zaniolo@cs.ucla.edu
}

\jdate{March 2000} \pubyear{2000}

\pagerange{\pageref{firstpage}--\pageref{lastpage}}

\begin{document}

\label{firstpage}

\maketitle

\noindent
\begin{abstract}
In the design of algorithms, the greedy paradigm
provides a powerful tool for solving efficiently classical
computational problems, within the framework of procedural
languages. However, expressing these algorithms
within the declarative framework of logic-based languages
has proven a difficult research challenge.
In this paper, we extend the framework of Datalog-like
languages to obtain simple declarative formulations
for such problems, and  propose effective
implementation techniques to ensure
computational complexities comparable
to those of procedural formulations. These
advances are achieved through
the use of the {\tt choice} construct,
extended with preference annotations to
effect the selection of alternative stable-models and
nondeterministic fixpoints.
We show that, with suitable storage
structures, the differential fixpoint computation of
our programs matches the complexity of
procedural algorithms in  classical search and
optimization problems.
\end{abstract}

\section{Introduction}
The problem of finding efficient implementations
for declarative logic-based languages represents one of the most arduous
and lasting research challenges in computer science.
The interesting theoretical challenges posed by this
problem are made more urgent by the fact that
extrema and other non-monotonic constructs
are needed to express many real-life applications,
ranging from the `Bill of Materials' to graph-computation
algorithms.

Significant progress in this area has been achieved on the
semantic front, where the introduction of the
well-founded model semantics and stable-model
semantics allows us to assign a formal meaning to
most, if not all, programs of practical interest.
Unfortunately, the computational problems remain
largely unsolved:  various approaches have been
proposed to more effective computations of  well-founded models and
stable models \cite{GelRos*91,gelf-lifs-88}, but
these fall far short of matching the
efficiency of classical procedural solutions for
say, algorithms that find shortest paths in graphs.
In general, it is known that determining whether a program has a
stable model is NP-complete \cite{MarTru91}.

Therefore, in this paper we propose a different approach:
while, at the semantic level, we strictly adhere to the
formal declarative semantics of logic programs with
negation, we also allow the use  of
extended non-monotonic constructs with first order semantics
to facilitate the task of programmers and
compilers alike. This entails simple declarative formulations
and nearly optimal executions for large classes of problems
that are normally solved using greedy algorithms.

Greedy algorithms~\cite{MorSha91}
are those that solve a class of optimization problems,
using a control structure of a single loop,
where at each iteration  some
element judged the `best' at that stage
is chosen and it is added to the
solution. The simple loop hints that these
problems are amenable to a fixpoint
computation.
The choice at each iteration calls
attention to mechanisms by which nondeterministic
choices can be expressed in logic programs.
This framework also provides an opportunity of
making, rather than blind choices, choices based
on some heuristic criterion, such as
greedily choosing the least (or most) among the  values at hand when
seeking the global minimization (or maximization)
of the sum of such values.
Following these hints, this paper introduces primitives for
choice and greedy selection, and
shows that classical greedy algorithms can be expressed using them.
The paper also shows how to translate
each program with such constructs to a program
which contains only negation as nonmonotonic construct,
and which defines the semantics of the original program.
Finally, several classes of programs with such
constructs are defined and it is shown that (i) they
have stable model semantics (ii) they are easily identifiable
at compile time, and (iii) they can be
optimized for efficient execution
---i.e., they yield the
same complexities as those expected from greedy
algorithms in procedural programs.
Thus, the approach provides a programmer with declarative tools
to express greedy algorithms, frees him/her from
many implementation details, yet guarantees
good performance.

Previous work has shown that many non deterministic decision
problem can be easily expressed using the nondeterministic
construct $choice$ in logic
programs~\cite{SacZan90,GiaPed*91}. In \cite{GPZ96}, we showed
that while the semantic of  choice requires the use of negation
under total stable model semantics, a stable model for these
programs can be computed in polynomial time. In fact,
choice in Datalog programs stratified with respect to negation
achieves DB-Ptime completeness under genericity \cite{AHV94}.
In this paper, we further explore the ability
of choice to express and support efficient computations,
by specializing choice with optimization heuristics expressed by
the $choice\-least$ and $choice\-most$ predicates.
Then, we show that these two new
built-in predicates enable us to
express easily greedy algorithms; furthermore,
by using appropriate data structures, the least-fixpoint
computation of a program with choice-least and choice-most
emulates the classical greedy algorithms, and achieves their
asymptoptic complexity.

A significant amount of excellent previous work has investigated
the issue of how to express in logic and compute
efficiently greedy algorithms, and, more in general, classical
algorithms  that require non-monotonic constructs. An
incomplete list include work by \cite{SudRam91,Die92,RosSag92,Gel93,GanGre*95}.  This line of research was often motivated by the
observation that many greedy algorithms can be viewed as
optimized versions of transitive closures.  Efficient computation
of transitive closures is central to deductive database
research, and the need for greedy algorithms is
pervasive in deductive database applications and in
more traditional database applications such as the
Bill of Materials~\cite{book97}.
In this paper, we introduce a treatment for greedy
algorithms that is significant simpler and
more robust  than previous approaches (including that proposed in
\cite{GreZan*92} where it was proposed to
use the $choice$ together with the built-in predicates $least$ and $most$); it also treats
all aspects of these algorithms, beginning from their intuitive formulation,
and ending with their optimized expression and execution.

The paper is organized as follows.
In Section 2 we present basic definitions on the syntax and
semantics of Datalog.
In Section 3, we
introduce the notion of choice and the stable-model
declarative semantics of choice programs.
In Section 4, we show how with  this non-deterministic construct
we can express in Datalog algorithms such as single-source
reachability and Hamiltonian path.
A fixpoint-based operational semantics for choice programs
presented in Section 5, and this semantics is then specialized
with the introduction of the choice-least and choice-most construct
to force greedy selections among alternative choices.
In Section 6, we show how the greedy refinement
allow us to express greedy algorithms such as Prim's and Dijskra's.
Finally, in  Section 7, we turn to the implementation of choice, choice-least
and choice-most programs, and show that using well-known  deductive DB
techniques, such as differential fixpoint,
and suitable access structures, such as hash tables and priority queues,
we achieve optimal complexity bounds for classical
search problems.

\section{Basic Notions}

In this section, we summarize the basic notions of Horn Clauses logic,
and its extensions to allow negative goals.

A {\em term} is a variable, a constant, or a complex term of the form
$f( t_1,\ldots, t_n )$, where $t_1,\ldots, t_ n$ are terms. An {\em
atom} is a formula of the language that is of the form $p( t_1,\ldots,
t_n)$ where $p$ is a predicate symbol of arity $n$.  A {\em literal }
is either an atom (positive literal) or its negation (negative
literal). A {\em rule} is a formula of the language of the form
\begin{center}
$ Q \leftarrow Q_1,\ldots,Q_m. $
\end{center}
where $Q$ is a atom ({\em head} of the rule) and $Q_1,\ldots,Q_m$ are
literals ({\em body} of the rule).  A term, atom, literal or rule is
{\em ground} if it is variable free. A ground rule with empty body is
a fact.  A logic program is a set of rules. A rule without negative
goals is called positive (a Horn clause); a program is called positive
when all its rules are positive. A DATALOG program is a positive
program not containing complex terms.

Let $P$ be a program. Given two predicate symbols $p$ and $q$ in
$P$, we say that $p$ {\em directly depends on} $q$, written $p \prec q$
if there exists a rule $r$ in $P$ such that $p$ is the head
predicate symbol of $r$ and $q$ occurs in the body of $r$.
The binary graph representing this relation is called
the {\em dependency graph} of $P$.
The maximal  strong components of this graph will
be called {\em recursive cliques}. Predicates in the
same recursive clique are {\em mutually recursive}.
A rule is {\em recursive} if its head predicate
symbol is mutually recursive with some
predicate symbol occurring in the body.

Given a logic program $P$, the Herbrand universe of $P$, denoted
$H_P$, is the set of all possible ground terms recursively constructed
by taking constants and function symbols occurring in $P$. The
Herbrand Base of $P$, denoted $B_P$, is the set of all possible ground
atoms whose predicate symbols occur in $P$ and whose arguments are
elements from the Herbrand universe. A {\em ground instance} of a rule
$r$ in $P$ is a rule obtained from $r$ by replacing every variable $X$
in $r$ by a ground term in $H_P$.  The set of ground instances of $r$
is denoted by $ground(r)$; accordingly, $ground(P)$ denotes
$\bigcup_{r \in P} ground (r)$.
A {\em (Herbrand) interpretation} $I$ of $P$
is any subset of $B_P$.
An model $M$ of $P$ is an interpretation that makes each ground instance
of each rule in $P$ $true$
(where a positive ground atom is $true$ if and only if it belongs to
$M$  and a negative ground atom is $true$ if and only if it
does not belong to $M$---total models).
A rule in $ground(P)$ whose body is true w.r.t. an interpretation $I$
will also be called fireable in $I$.
Thus, a model for a program can be constructed by a procedure
that starts from $I:= \emptyset$ and adds to $I$ the head
of a rule $r \in ground(P)$ that is firable in $I$ (this operation
will be called firing $r$) until no firable rules remain.
A model of $P$ is {\em minimal} if none of its proper subsets
is a model.  Each positive logic program has a unique minimal model
which defines its formal declarative semantics.

Given a program $P$ and an interpretation $M$ for $P$,
we denote as $ground_M ( P )$ the
program obtained from $ground(P)$ by

\begin{enumerate}
\item
removing every rule having as a goals some literal $\neg q$ with $q \in M$
\item
removing all negated goals from the remaining rules.
\end{enumerate}

Since $ground_M (P)$ is a positive program, it has a unique minimal model.
A model $M$ of $P$ is said to be {\em stable}
when $M$ is also the minimum model of
$ground_M (P)$ \cite{gelf-lifs-88}.  A given program can have
one or more stable (total) model, or possibly none.
Positive programs, stratified programs~\cite{AptBla*88},
locally stratified programs~\cite{Prz88} and weakly
stratified programs~\cite{PrzPrz88}
are among those that  have exactly one stable model.

Let $I$ be an interpretation for a program $P$.
The {\em immediate consequence operator} $T_P (I)$ is defined
as the set containing the heads of each rule $r \in ground(P)$ s.t. all
positive goals of $r$ are in $I$, and none of the negated goals of $r$,
is in $I$.


\section{Nondeterministic Reasoning}

Say that
our university database contains
a relation $\tt student(Name,  Major, Year)$,
and a relation $\tt professor(Name, Major)$.
In fact, let us take a toy example that only has
the following facts:

\bldl
\pfact{ student('Jim Black', ee, senior). \hspace*{1 cm} } { professor(ohm, ee). }
\pfact{ } { professor(bell, ee). }
\eldl

Now, the rule is that the major of a student must match his/her
advisor's major area of specialization.
Then eligible advisors can be computed as
follows:

\bldl
\prule{elig\_adv(S, P) } { student(S,  Majr, Year),\ professor(P, Majr).}
\eldl

This yields

\bldl
\pfact{elig\_adv('Jim Black', ohm).} {}
\pfact{elig\_adv('Jim Black', bell).} {}
\eldl

But, since a  student can only have one advisor,
the goal $\tt choice((S),(P))$ must be added to
force the selection  of a unique advisor, out of
the eligible advisors, for a student.

\begin{example} {\em Computation of unique advisors by choice rules}
\label{advex}
\bldl
\prule{ actual\_adv(S, P) } {student(S, Majr, Yr),\ professor(P, Majr),}
\pbody{}      {choice((S),(P)).}
\eldl
The computation of this rule gives for each student $\tt S$ a unique
professor $\tt P$ 
\end{example}

The goal $\tt choice((S),(P))$ can also be viewed as
enforcing a {\em functional dependency} (FD)  ${\tt S} \rightarrow {\tt P}$
on the results produced by the rule;
thus, in $\tt actual\_adv$, the second column (professor name) is
functionally dependent on the first one (student
name).

The result of executing this rule is {\em nondeterministic}.
It can either give a singleton relation containing
the tuple $\tt ('Jim Black', ohm)$ or that
containing the tuple $\tt ('Jim Black', bell)$.

A program where the rules contain choice goals is called
a {\em choice program}.
The semantics of a choice program $P$ can be defined
by transforming $P$ into a program with negation,
$foe(P)$, called  the {\em first order equivalent} of a choice program $P$.
$foe(P)$ exhibits a multiplicity of stable models,
each obeying the FDs defined
by the choice goals.
Each stable model for $foe(P)$ corresponds to an
alternative set of answers for  $P$ and
is called a {\em choice model} \ for $P$.
$foe(P)$  is defined as follows:

\begin{definition}\label{sv}\cite{SacZan90}
The first order equivalent version $foe(P)$ of a choice program $P$ is
obtained by the following transformation. Consider
a choice rule $r$ in $P$:
\[
r : A \leftarrow B(Z),\ choice((X_1), (Y_1)),\ \ldots,\ choice((X_k), (Y_k)).
\]
where
\begin{enumerate}
\item [(i)] $B(Z)$ denotes the conjunction of all the goals of $r$ that
are not choice goals, and
\item [(ii)] $X_i , \ Y_i , \ Z$, $1 \leq i \leq k$,
denote vectors of variables occurring in the body of $r$ such that
$X_i \cap Y_i = \emptyset$ and $X_i,Y_i \subseteq Z$.
\end{enumerate}
Then, $foe(P)$ is constructed by transforming
the original program $P$ as follows:

\begin{enumerate}
\item
Replace $r$ with a rule \ $r'$ obtained
by substituting the choice goals with the atom  $chosen_r(W)$:
\[
r' : A \leftarrow B(Z), \ chosen_r(W) .
\]
where $W \subseteq Z$ is the list of all variables appearing in choice
goals, i.e., $W = \bigcup_{1 \leq j \leq k} X_j \cup Y_j$.
\vspace*{2mm}
\item Add the new rule
\[ chosen_r (W) \leftarrow B(Z), \ \neg diffchoice_r (W). \]
\item
For each choice atom $choice((X_i),(Y_i))$ ($1 \leq i \leq k$),
add the new rule
\[
diffchoice_r (W) \leftarrow chosen_r(W'),  \ Y_i \neq Y'_i.
\]
where (i) the list of variables $W'$ is derived from $W$ by replacing
each $A \not\in X_i$ with a new variable $A'$ (i.e., by priming those variables),
and (ii) $Y_i \neq Y '_i$ is true if $A \neq A '$, for some
variable  $A \in Y_i$ and its primed counterpart $A' \in Y' _i$. 
\end{enumerate}
\end{definition}

The first order equivalent version  of Example~\ref{advex} is given in
Example~\ref{order}, which can be read as a statement that
a professor will be assigned
to a student whenever
a different professor has not been assigned to
the same student.

\begin{example} {The first order equivalent version of the rule in Example~\ref{advex}}
\label{order}
\bldl
\prule{actual\_adv(S,P) } { student(S, Majr, Yr),\ professor(P, Majr),}
\pbody{                 } { chosen(S,P).}
\prule{chosen(S, P)     } { student(S, Majr, Yr),\ professor(P, Majr),}
\pbody{                 } { \neg diffchoice(S,P).}
\prule{ diffchoice(S,P) } { chosen(S,P'),\ P \neq P'.}
\eldl
\end{example}

In general, the program $foe(P)$ generated by
the transformation discussed above
has the following properties\cite{GiaPed*91}:

\begin{itemize}
\item
$foe(P)$ has one or more total stable models.
\item
The {\em chosen} atoms in each stable model of $foe(P)$
obey the FDs defined by the choice goals.
\end{itemize}
The stable models of $foe(P)$ are called {\em choice models} for $P$.

While the topic of operational semantics for
choice Datalog programs will be further discussed
in Section \ref{fixpoint}, it is clear that choice
programs can be implemented efficiently. Basically, the
{\em chosen} atoms must be produced  one-at-a-time
and memorized in a table.
The {\em diffchoice} atoms need not be computed
and stored ({\em diffchoice} rules are not range restricted
and their evaluation could produce huge results);
rather, a goal {\em $\neg$diffchoice(t)} can simply
be checked dynamically against the table {\em chosen}.
Since these are simple operations (actually quasi constant-time
if an hash table is used), it follows that
choice Datalog programs can be computed in polynomial time,
and that rules with choice
can be evaluated as efficiently as those without choice.


\section{Computing with Choice} \label{choicealgos}

Choice significantly extends the power of Datalog, and Datalog
with stratified negation \cite{GreSac*95,GiaPed*91}.
In this paper we consider Datalog with the nondeterministc
construct $choice$, although our framework can be easily extended
to also consider stratified negation.

The following example presents a choice program that
pairwise chains the elements of
a relation $\tt d(X)$, thus establishing a random total order on
these elements.

\begin{example}\label{linear1}
{\em Linear sequencing of the elements of a set.} The
elements of the set are stored by means of facts of the form $\tt d(Y)$.
\bldl
\pfact{ succ(root,root). } { }
\prule{ succ(X,Y) } { succ(\_,X), \ d(Y),}
\pbody{} {choice((X),(Y)), \ choice((Y),(X)). \ \ \ \ }
\eldl
\end{example}

Here  $\tt succ(root,root)$ is the root of a chain
linking all the elements of $\tt d(Y)$. The transitive closure
of $\tt succ$ thus defines a total order on the elements of $\tt d$.
Because of the ability of choice programs to order the
elements of a set,
Datalog with choice is P-time complete and can,
for instance, express the parity query---i.e.,
determining
if a relation has an even number of elements~\cite{AHV94}.
This query cannot be expressed in  Datalog with stratified negation
unless we assume that the underlying universe
is totally ordered---an assumption that
violates the data independence principle of
{\em genericity} \cite{ChH82,AHV94}.

The expressive power of the choice
construct has been studied in \cite{GPZ96,GreSac*95}, where it
is shown that  it is  more powerful
than other nondeterministic constructs, including
the witness operator~\cite{AbiVia91},
and the original version of choice  proposed in \cite{KriNaq88},
which is called static-choice, to distinguish it from the dynamic choice
used here \cite{GiaPed*91}. For instance, it has been shown
in \cite{GiaPed*91}, that the task  of ordering a domain
or computing whether a relation contains an even number of elements
(parity query) cannot be performed by positive programs
with static choice or the witness operator \cite{AbiVia91}.

In the rest of the paper, we will study nondeterministic
queries combined with optimization criteria.
For instance, our previous advisor example can be
modified using optimized criteria to match students with
candidate advisors.
In the next example we present the general matching problem for
bipartite graphs.

\begin{example}\label{matching-ex1}
{\em Matching in a bipartite graph.}
We are given a bipartite graph $G = \< (V_1,V_2), E \>$, i.e.
a graph where nodes are partitioned into two subset
$V_1$ and $V_2$ and each edge connect nodes in $V_1$ with nodes in $V_2$.
The problem consists to find a matching, i.e., a subset $E'$ of $E$ such
that each node in $V_1$ is joined with at most one edge in $E'$ with a
node in $V_2$ and vice versa.
\bldl
\prule{ matching(X,Y) } { g(X,Y,C), choice((Y),(X)). }
\pbody{                } { choice((X),(Y)), choice((X),(C)). }
\eldl
Here a fact $g(x,y,c)$ denotes that there is an edge with cost $c$
joining the node $x \in V_1$ with the node $y \in V_2$.
\end{example}

In section \ref{leastalgos}, we will consider the
related optimization problem, of finding a matching such
that the sum of all $\tt C$s is
minimized or maximized\footnote{Given that the
pair $\tt X \rightarrow Y$, $\tt X \rightarrow C$
is equivalent to $\tt X \rightarrow Y, C$, the last rule in
the previous example can  also
be written as follows:

\vspace{2mm}
\hspace{1cm}
$\tt matching(X,Y) \leftarrow g(X,Y,C), choice((Y),(X)), choice((X),(Y,C)). $
}.

\begin{example}\label{span1}
{\em Rooted spanning tree.} We are given an
undirected graph where an
edge joining two nodes, say $x$ and $y$, is represented by means
of two facts $g(x,y,c)$ and $g(y,x, c)$, where $c$ is the cost.
A spanning tree in the graph, starting from the source node $\tt a$,
can be expressed by means of the following program:

\vspace{-2mm}
\bldl
\pfact{ st(root,a, 0). } { }
\prule{ st(X,Y, C)        } { st(\_,X, \_), \ g(X,Y, C), \ Y \neq a, \ Y \neq X, }
\pbody{}{choice((Y),(X)), choice((Y), (C)). }
\eldl
\noindent
To illustrate the presence of
multiple total choice models for this program,
take a simple graph $G$ consisting of the following arcs:
\bldl
\pfact{ g(a, b, 1). \ \ \ \ \ } { g(b, a, 1).}
\pfact{ g(b, c, 2). } { g(c, b,  2).}
\pfact{ g(a, c, 3). } { g(c, a,  3).}
\eldl

After the exit rule adds $\tt st(root, a, 0)$,
the recursive rule could add  $\tt st(a, b, 1)$ and
$\tt  st(a, c, 3)$ along with the two tuples
$\tt  chosen(a,b,1)$ and $\tt chosen(a,c,3)$
in the $\tt chosen$ table.
No further arc can be added after those,
since the addition of $\tt st(b, c, 2)$ or
$\tt st(c, b, 2)$ would violate
the FD that follows from $\tt choice((Y),(X))$
enforced through the $\tt chosen$ table.
However, since
$\tt st(root,$ $\tt a, 0)$, was produced by the first rule (the exit rule),
rather than the second rule (the recursive choice rule), the
table $\tt chosen$ contains no tuple with second argument equal to the
source node $\tt a$.
Therefore, to avoid the addition of
$\tt st(c, a, 3)$ or $\tt st(b, a, 1)$, the goal
$\tt Y \neq a$ was added to the recursive rule.

By examining all possible solutions, we
conclude that this program has three different choice models:
$M_1 = \{ {\tt st(a, b, 1), st(b, c, 2)} \} \cup G$,
$M_2 = \{ {\tt st(a, b, 1), st(a, c, 3)} \} \cup G$ and
$M_3 = \{ {\tt st(a, c, 3), st(c, b, 2)} \} \cup G$.
\end{example}

\begin{example}\label{reachability}
{\em Single-Source Reachability.}
Given a direct graph where the arcs are stored by means of tuples
of the form $g(x,y,c)$, the set of nodes reachable from a node $\tt a$
can be defined by the following program:
\bldl
\pfact{ reach(a,0). } { }
\prule{ reach(Y,C)  } { reach(X,C_1),\ g(X,Y,C_2),\ Y \neq a,}
\pbody{}{C=C_1+C_2,\ choice((Y),(C)). }
\eldl
\end{example}

Once the cost arguments are eliminated from these rules,
we obtain the usual transitive-closure-like program,
for which the fixpoint  computation terminates once all
nodes reachable from node $\tt a$ are found, even if the
graph contains cycles. However,
if the choice goal were eliminated,
the program of Example \ref{reachability}
could become nonterminating on a cyclic graph.

In the next example, we have a complete undirected
labeled graph $G$,
represented by facts $ g(x, y, c)$, where the label
$\tt c$ typically represents the cost of the edge.
A simple path is a path passing through a node
at most once.
A Hamiltonian path is a simple path reaching each node in the graph.
Then, a simple path can be constructed as
follows:

\begin{example}\label{hamiltonian} {\em The simple path problem.}
When the arc from $\tt X$ to $\tt Y$  is selected, we must
make sure that the ending node $\tt Y$ had not been
selected and the starting node $\tt X$ is connected to some
selected node. The choice constraints, and the goals
$\tt  s\-path(root, Z, 0),  Y \neq Z$
to avoid returning to the initial node, ensure that
a simple path is obtained.
\bldl
\hspace*{-5mm}
\prule{ s\-path(root,X,0) } { g(X,\_,\_), \ choice((),X)).}
\hspace*{-5mm}
\prule{ s\-path(X,Y,C)    } { s\-path(\_,X,\_),\ g(X,Y,C), \ s\-path(root, Z, 0), \ Y \neq Z, }
\pbody{                   } { choice((X),(Y)),\ choice((Y), (X)), \
                              choice((Y), (C)). }
\eldl
\end{example}

When $G$ is  a complete graph,
the simple path produced by this program
is Hamiltonian (i.e., touches all the nodes).
In many applications, we need to find a minimum-cost Hamiltonian
path; this is the Traveling Salesman Problem (TSP) discussed
in Section \ref{leastalgos}.

The next program presents a problem consisting in the selection of a set of elements
satisfying a constraint.
The optimized version of this problem is the well-known knapsack problem.


\section{Fixpoint Semantics} \label{fixpoint}

\subsection{Choice programs}

Let $I$ be an interpretation for a program $P$;
the {\em immediate consequence operator} $T_P (I)$ is defined
as the set containing the heads of each rule $r \in ground(P)$ s.t. all
positive goals of $r$ are in $I$, and none of the negated goals of $r$,
is in $I$.
For a choice program $P$, with first order equivalent
$foe(P)$, let us denote by $T_{P_{C}}$ the
immediate consequence operator associated with the
rules defining the predicate $\tt chosen$
in $foe(P)$
(these are the rules with the $\tt \neg diffchoice$ goals)
and let $T_{P_D}$ denote the immediate consequence
for all the other  rules in $foe(P)$ (for positive choice
programs these are Horn clauses).

Therefore,
we have that, for any interpretation $I$ of $foe(P)$:
\[
T_{foe(P)}(I) = T_{P_D}(I) \ \cup \ T_{P_C}(I).
\]

Following \cite{GPZ96} we can now introduce a general operator for
computing the nondeterministic fixpoints of a choice program $P$.
We will denote by
$FD_P$ the functional dependencies defined by
the choice goals in $P$.

\begin{definition}\label{psi}
Given a choice program $P$, its {\em nondeterministic
immediate consequence operator} $\Psi_P$ is a
mapping from an interpretation of $foe(P)$ to a set
of interpretations of $foe(P)$ defined as follows:
\begin{equation}
\hspace*{-5mm}
\Psi_P(I) = \{\  T_{P_D}^{\uparrow \omega}(I \cup \Delta C) \cup \Delta C
                \mid\ \Delta C \in \Gamma_P(I)\ \}
\end{equation}
where: $\Gamma_P(I) = \{ \emptyset \}$ if $T_{P_{C}}(I)= \emptyset$, and
otherwise:
\begin{equation}
\label{psi2}
\begin{array}{lll}
\hspace*{-5mm}
\Gamma_P(I) & = & \{ \Delta C \ \mid \ \emptyset \subset \Delta C
\subseteq T_{P_{C}}(I)
           \setminus I \ {\rm and} \
           I \cup \Delta C \models FD_P \ \}
\end{array}
\end{equation}

\noindent
with $I \cup \Delta C \models FD_P$
denoting that $I \cup \Delta C$ satisfies the dependencies in $FD_P$.
\end{definition}

\vspace{0.2cm}
\noindent
Therefore, the $\Psi_P$ operator is basically the composition
of two operators. Given an interpretation
$I$, the  first operator computes all
the admissible subsets of $\Delta C \subseteq T_{P_C} (I)$, i.e., those
where $I \cup \Delta C$ obeys the given FDs;
the second operator derives the logical consequence for each
admissible subset using the $\omega$-power of $T_{P_D}$.

The definition of $\Gamma_P(I)$ is such that
$\Delta C$ is not empty iff $T_{P_C}(I) \setminus I$
is not empty; thus, if there are possible new choices, then at least
one has to be taken.
The $\Psi_P$ operator formalizes a single step of a bottom-up
computation of a choice program. Instead of defining the powers of
$\Psi_P$, it is technically more convenient to define directly the
notion of a nondeterministic computation based on the $\Psi_P$
operator.

Observe that given the presence of the constraint,
$I \cup \Delta C \models FD_P$,
we can eliminate the $\tt \neg diffchoice$
goal from the chosen rules.
In fact, if $T_{P'_{C}}$ denotes the immediate consequence operator for
the chosen rules without the $\tt \neg diffchoice$ goals,
then $T_{P'_{C}}$ can replace
$T_{P_{C}}$ in Equation \ref{psi2}.

\begin{definition}\label{def:comput}
{\em
Given a choice program $P$,
an {\em inflationary choice fixpoint computation} for $P$, is
a sequence $\langle I_n \rangle_{n \geq 0}$ of interpretations
such that:
\begin{itemize}
\item [i.]  $I_0 = \emptyset$,
\item [ii.] $I_{n+1} \in \Psi_P(I_n)$, \quad for $n \geq 0$.
\end{itemize}
}
\end{definition}

Inasmuch as every sequence $\langle I_n \rangle_{n \geq 0}$ is monotonic,
it has a unique limit for $n \rightarrow \infty$; this limit will
be called an {\em inflationary choice fixpoint} for the choice
program $P$. Thus, we have the following result:

\begin{theorem} \cite{GiaPed*91}
Let $P$ be a Datalog program with choice,
and $M$ a Herbrand interpretation for $foe(P)$.
Then $M$ is a choice model for $P$ iff $M$ is an
inflationary choice fixpoint for $P$. 
\end{theorem}

Moreover, the inflationary choice fixpoint is {\em sound} (every result
is a choice model) and {\em complete} (for each choice model there is some
inflationary choice fixpoint computation producing it).
For logic programs with infinite Herbrand universe,
an additional assumption of
{\em fairness} is needed to ensure completeness \cite{GPZ96}.
As customary for database queries, computational
complexity is evaluated with respect to the size
of the database. Then, we have the following result:

\begin{theorem}  \cite{GiaPed*91}
Let $P$ be a choice Datalog program.
Then, the data complexity of
computing a choice model for $P$ is polynomial time.
\end{theorem}

Therefore, for a choice Datalog program, $P$,
the computation of one of the stable models for $foe(P)$
can be performed in polynomial time using the Choice Fixpoint Computation.
This contrasts with the general intractability of
finding stable models for general programs: in
fact, we know that checking if a Datalog  program with negation  has a
stable model is NP-complete \cite{MarTru91}.

Therefore, the choice construct allows us to capture a special
subclass of programs that
have a stable model semantics but are
amenable to efficient implementation and are
appealing to intuition. Implementing these
programs only requires memorization of the $chosen$ predicates;
from these, the {\em diffchoice} predicates can be generated on-the-fly,
thus eliminating the need to store {\em diffchoice}
explicitly. Moreover, the model of memorizing
tables to enforce functional dependencies provides
a simple enough metaphor for a programmer to make effective
usage of this construct without having to become cognizant on
the subtleties of non-monotonic semantics.
We conclude by mentioning that, although we are considering
(positive) choice Datalog programs, our framework can be
trivially extended to also consider stratified negation.
The computation of a choice model for a stratified choice
program can be carried out by partioning the program into
an ordered number of suitable subprograms (called 'strata') and  computing
the choice fixpoints of every stratum in their order.

\subsection{Greedy Choice}

Definition \ref{psi} leaves quite a bit of
latitude in the computation of $\Delta$ (Equation \ref{psi2}).
This freedom can be used to select $\Delta$s that have
additional properties. In particular, we want to
explore specializations of this concept
that trade nondeterministic completeness (which
is only of abstract interest to a programmer) in return
for very concrete benefits, such as expressive
power and performance.
For instance, in the specialization called
{\em Eager Choice}~\cite{GiaPed*91}, a maximal
$\Delta C$ is used in Equation \ref{psi2}.
This results in a significant increase in expressive power,
as demonstrated by the fact that negation can be emulated by
eager choice ~\cite{GiaPed*91,GPZ96}.

In this paper, we focus on a specialization of choice called
{\em greedy choice}; our interest in this constructs
follows from the observation
that it is frequently desirable
to select a  value that is the {\em least}
(or the {\em most}) among the possible values and still
satisfy the FDs defined by the choice atoms.

A choice-least (resp. choice-most) atom is of the form
{\tt choice\-least((X),(C))} (resp. {\tt choice\-most((X),(C))} )
where $\tt X$ is a list of variables and $\tt C$ is a single variable
ranging over an ordered domain.
A rule may have at most one choice-least or one choice-most atom.
A goal {\tt choice\-least((X),(C))} (resp. {\tt choice\-most((X),(C))}) in a rule $r$
can be used to denote that  the FD defined by the atom {\tt choice((X),(C))} is to be satisfied ---
the declarative semantics of choice, choice-least and choice-most coincide.
For instance,  a rule of the form
\bldl
\prule{ p(X,Y,C) } { q(X,Y,C), choice((X),(Y)), choice\-least((X),(C)). }
\eldl
defines the FD $\tt X \rightarrow Y,C$ on the possible instances of $\tt p$.
Thus, assuming that $\tt q$ is defined by the facts $q(a,b,1)$ and $q(a,c,2)$,
from the above rule we can derive either $p(a,b,1)$ or $p(a,c,2)$.
Moreover, the choice-least goal introduces some heuristic in the computation
to derive only $p(a,b,1)$.
This means that, by using choice-least and choice-most predicates, we introduce
some preference criteria on the stable models of the program.
The `greedy' fixpoint computation permit us to compute a `preferred' stable model.

We can now define a {\em choice-least rule} (resp.  {\em choice-most rule})
as one that contains one choice-least (resp. one choice-most) goal,
and zero or more choice goals.
Moreover, we also assume that our programs contain either choice-least or
choice-most rules.
A program that contains choice-least rules (choice-most rules)
and possibly other rules with zero or more choice goals
is called a {\em choice-least} program (a choice-most program) .
Choice-least and choice-most programs have dual properties;
thus in the rest of the paper we will often mention the
properties of one kind of program with the understanding
that the corresponding properties of the other  are implicitly
defined by this duality.

The correct computation of {\em choice-least} programs
can be thus defined by specializing the nondeterministic immediate
consequence operator by (i) ensuring that $\Delta$ is a
singleton set, containing only one element (ii) ensuring that a least-cost
tuple among those that are candidates is chosen.

Formally, we can use as our starting point
the {\em lazy version of choice} where $\Delta$ is specialized
into a singleton set $\delta$. The specialized version of
$\Psi_P$ so derived will be denoted $\Psi_P^{lazy}$; as
proven in \cite{GiaPed*91}, the inflationary choice fixpoint
restricted using $\Psi_P^{lazy}$ operators still provides
a sound and nondeterministically
complete  computation for the choice models of $P$.

We begin by decomposing $\Psi_P^{lazy}$ in three steps:

\begin{definition}\label{LICO-def}
{\em Lazy Immediate-Consequence Operator (LICO).}

Let $P$ be a choice program and $I$ an interpretation of $P$. Then
$\Psi_{P} (I)$  for $P$ is defined as follows:
\[
\begin{array}{lll}

\Theta_I    & = & \{ \delta \in  T_{P_{C}}(I) \setminus  I  \ \mid \
                  I \cup \{ \delta  \} \models FD_P \} \\
\Gamma_P^{lazy}(I) & = & \{ I \cup \{ \delta \} \
                  \mid \ \delta \in \Theta_I \}
                  \ \cup \ \{ I \ \mid \  \Theta_I = \emptyset \} \\
\Psi_P^{lazy}(I)  & =  & \{ \  T_{P_D}^{\uparrow \omega}(J) \mid\ J
\in \Gamma_P^{lazy}(I)\ \}
\end{array}
\]
\end{definition}

Given an interpretation $I$, a set $\Delta \in \Gamma_P(I)$
and two tuples $t_1, t_2 \in \Delta$.
We say that $t_1 < t_2$ if both tuples are inferred only by
choice-least rules and the cost of $t_1$ is lesser than the cost of $t_2$.
Further, we denote with $least(\Delta)$ the set of tuples of $\Delta$
with least cost, i.e.
$least(\Delta) = \{ t | t \in \Delta \mbox{ and } \not\exists u \in \Delta \mbox{ s.t. }
u < t \}$.

Therefore, the implementation of greedy algorithms follows directly from
replacing $\delta \in \Theta_I$ with
$\delta \in least (\Theta_I)$.

\newpage
\begin{definition}\label{greed}
{\em Least-Cost Immediate-Consequence Operator.}

Let $P$ be a choice program and $I$ an interpretation of $P$. Then
$\Psi_P^{least} (I)$  for $P$ is defined as follows:
\[
\begin{array}{lll}
\Theta_I    & = & \{ \delta \in  T_{P_{C}}(I) \setminus  I  \ \mid \
                  I \cup \{ \delta  \} \models FD_P \} \\

\Gamma_P^{least}(I) & = & \{ I \cup \{ \delta \} \
                  \mid \ \delta \in least (\Theta_I) \}
                  \ \cup \ \{ I \ \mid \  \Theta_I = \emptyset \}\\

\Psi_P^{least}(I)  & =  & \{ \  T_{P_D}^{\uparrow \omega}(J) \mid\ J
\in \Gamma_P^{least}(I)\ \}

\end{array}
\]

\noindent
$\Psi_P^{least}$  will be called the
{\em Least-Cost Immediate-Consequence Operator}. 
\end{definition}

Likewise, we have the dual definition of
the {\em Most-Cost Immediate-Consequence Operator}.

\begin{definition}
Let $P$ be a program with choice and choice-least goals.
An {\em inflationary least choice fixpoint computation} (LFC) for $P$, is
a sequence $\langle I_n \rangle_{n \geq 0}$ of interpretations
such that:
\begin{itemize}
\item[i.]  $I_0 = \emptyset$,
\item[ii.] $I_{n+1} \in \Psi_P^{least}(I_n)$, \quad for $n \geq 0$.
\end{itemize}
\end{definition}

Thus, all the tuples that do not violate the given FDs (including
the FDs implied by least) are considered, and one is chosen
that has the least value for the cost argument.

\begin{theorem}
Let $P$ be a Datalog program with choice and choice\_least.
Then,
\begin{enumerate}
\item
every inflationary least choice fixpoint for  $P$ is a choice model for $P$.
\item
every inflationary least choice fixpoint of $P$ can be computed
in polynomial time.
\end{enumerate}
\end{theorem}

\noindent
\begin{proof}
For the first property, observe that
every computation of the inflationary least choice fixpoint
is also a compuation of  the lazy choice fixpoint. Therefore
every inflationary least choice fixpoint for  $P$ is a choice model for $P$.

The second property follows from the fact that
the complexity of the inflationary lazy choice fixpoint  is polynomial
time. Moreover, the cost of selecting a tuple with least cost is also
polynomial. Therefore, the complexity of inflationary least choice fixpoint
is also polynomial.
\end{proof}

While the inflationary choice fixpoint computation is
sound and complete with respect to the declarative stable-model
semantics the  inflationary least (most) choice fixpoint computation
is sound but no longer complete; thus there are choice models
that are never produced by this computation. Indeed,
rather than following
a ``don't care" policy when choosing among stable models, we
make greedy selections between the available alternatives.
For many problems of interest, this greedy
policy is sufficient to ensure that the resulting
models have some important optimality properties, such as the
minimality of the sum of cost of the edges.
The model so constructed, will be called
{\em greedy choice models}\footnote{In terms of relation
between declarative and operational semantics, the situation
is similar to that of pure Prolog programs, where the the
declarative semantics is defined by the set of all legal SLD-trees,
but then one particular tree will be generated instead of others
according to some preference criterion}.

\section{Greedy Algorithms} \label{leastalgos}
In a system that adopts  a concrete
semantics based on {\em least choice fixpoint}, a programmer will
specify a {\tt choice-least((X),(Y))} goal
to ensure that only particular
choice models rather than arbitrary ones
are produced, through the greedy selection of the
least values of $\tt Y$  at each step.   Thus an optimal
matching in a directed graph problem can be expressed as follows:

\begin{example}\label{minimum-matching-ex}
{\em Optimal Matching in a bipartite graph}
\bldl
\prule{opt\_matching(X,Y)} {g(X,Y,C),\ choice((Y),(X)), }
\pbody{                } { choice((X),(Y)),\ choice\-least((X),(C)). }
\eldl
\end{example}

Observe that this program is basically that of Example \ref{matching-ex1}
after that the choice goal with a cost argument
has been specialized to a choice-least  goal.

The specialization of choice goals into  {\em choice-least}
or {\em choice-most} goals yields
a convenient and efficient formulation of many
greedy algorithms, such as Dijkstra's shortest path
and Prim's minimum-spanning tree algorithms
discussed next.

The algorithm for finding the  minimum spanning tree
in a weighted graph, starting from a
source node $\tt a$,
can  be derived from the program of Example~\ref{span1} by
simply replacing the goal $\tt choice((Y),(C))$ with
$\tt choice\-least((Y),$ $\tt (C))$ yielding the well-known Prim's
algorithm.

\begin{example}\label{prim1}
{\em Prim's Algorithm.}

\vspace{-2mm}
\bldl
\pfact{ st(root,a, 0). } { }
\prule{ st(X,Y,C)      } { st(\_,X,\_), \ g(X,Y, C), \ Y \neq a, }
\pbody{                } { choice((Y),(X)), \ choice\-least((Y), (C)). }
\eldl
\end{example}

Analogously, the algorithm for finding the shortest path
in a weighted digraph, starting from a
source node $\tt a$,
can  be derived from the program of Example~\ref{reachability} by
simply replacing the goal $\tt choice((Y),(C))$ with
$\tt choice\-least((Y),$ $\tt (C))$,
yielding the well-known Dijkstra's
algorithm, below.

\begin{example}\label{dijkstra}
{\em Dijkstra's algorithm.}
\vspace{-2mm}
\bldl
\pfact{ dj(a,0). } { }
\prule{ dj(Y,C)  } { dj(X,C_1),\ g(X,Y,C_2),\ Y \neq a, }
\pbody{}            {C=C_1+C_2,\  choice\-least((Y),(C)). }
\eldl
\end{example}

Consider now the program of  Example \ref{linear1},
which chains the elements of a
domain $\tt d(X)$ in an arbitrary order.
Say now that a particular
lexicographical order is pre-defined and we
would like to sort the elements of $\tt d(X)$
accordingly.
Then, we can write the rules as follows:

\begin{example}\label{sorting} {\em Sequencing the elements
of a relation in decreasing order.}
\bldl
\pfact{ succ(root,root).} { }
\prule{ succ(X,Y)       } { succ(\_,X),\ d(Y),}
\pbody{                 } { choice\-most((X),(Y)), \ choice((Y),(X)). }
\eldl
\end{example}

Greedy algorithms often provide efficient approximate
solutions to NP-complete problems; the following algorithm
yields heuristically effective approximations of optimal solutions
for the traveling salesperson problem \cite{PapLew75}.

\begin{example}\label{tsp} {\em Greedy TSP.}

Given a complete undirected graph, the
exit rule simply selects an arbitrary node $\tt X$,
from which to start the search. Then, the recursive rule
greedily chooses at each step an arc $\tt (X,Y,C)$ of least
cost $\tt C$ having $\tt X$ as
its end node.
\bldl
\prule{ s\-path(root,X,0)} { node(X), \ choice((),X).}
\prule{ s\-path(X,Y,C)     } { s\-path(\_,X,\_),\ g(X,Y,C),}
\pbody{}                     {s\-path(root, Z, 0), \ Y \neq Z, }
\pbody{                    } { choice((X),(Y)), \  choice((Y),(X)),}
\pbody{}                     { choice\-least((Y),(C)). }
\eldl
\end{example}

Observe that the program of Example \ref{tsp} was obtained
from that of Example \ref{hamiltonian} by replacing a choice
goal with its choice-least counterpart.

\begin{example}\label{Knapsack}
While we have here concentrated on graph optimization problems,
greedy algorithms are useful in a variety of other problems.
For instance, in \cite{report}
it is presented a greedy solution to the
well-known knapsack problem consists in finding a set
of items whose total weight is lesser than a given value (say 100)
and whose cost is maximum.
This is an NP-complete problem and, therefore, the optimal solution
requires an exponential
time (assuming $P \neq NP$) but an approximate solution
carried out by means of a greedy computation,
which selects at each step the item with maximum
$value/weight$ ratio.
\end{example}

In conclusion, we have obtained
a framework for deriving  and expressing greedy algorithms
(such as Prim's algorithm) characterized by conceptual simplicity,
logic-based semantics, and short and efficient programs;
we can next turn to the efficient implementation problem for
our programs.


\section{Implementation and Complexity}\label{algorithms}

A most interesting aspect of the
programs discussed in this paper is that their
stable models can be computed
very efficiently.  In the previous sections, we have seen that
the exponential intractability of stable
models is not an issue here: our greedy
fixpoint computations are always polynomial-time in
the size of the database. In this section, we show that
the same asymptotic complexity
obtainable by expressing  the algorithms in
procedural languages can be obtained  by using
comparable data structures and taking advantage of
syntactic structure of the program.

In general, the computation consists
of two phases: (i) compilation and (ii) execution.
All compilation algorithms discussed here
execute with time complexity that is polynomial
in the size of the programs.
Moreover, we will assume, as it is customarily done \cite{book97},
that the size of the database dominates that of the
program. Thus, execution costs dominate the compilation costs,
which can thus be disregarded in the derivation of
the worst case complexities.

We will use compilation techniques, such as the
differential fixpoint computation, that are of common usage in
deductive database systems \cite{book97}.
Also we will employ suitable storage structures, such as
hash tables to support search on keys, and
priority queues to support choice-least and choice-most goals.

We assume that our programs consist of a set of mutually recursive
predicates. General programs can be partitioned into a set
of subprograms where rules in every subprogram defines a set of mutually
recursive predicates. Then, subprograms are computed
according to the topological order defined by the
dependencies among predicates, where tuples derived from the computation
of a subprogram are used as database facts in the computation of the
subprograms that follow in the topological order.

\subsection{Implementation of Programs with Choice}\label{Choice-imple}

Basically, the {\em chosen} atoms need to be
memorized in a set of tables ${\tt chosen}_r$ (one for each
${\tt chosen}_r$ predicate).
The {\em diffchoice} atoms need not be computed
and stored; rather, a goal {\em $\neg$diffchoice}$_r( \ldots )$ can simply
be checked dynamically against the table ${\tt chosen}_r$.
We now present how programs with choice can be evaluated by means of
an example.

\begin{example} \label{exp11}
Consider again Example \ref{sorting}

\bldl
s_1:\pfact{succ(root,root).} { }
s_2:\prule{ succ(X,Y)       } { succ(\_,X),\ d(Y),}
\pbody{                 } { choice\-most((X),(Y)), \ choice((Y),(X)). }
\eldl
\end{example}

According to our definitions, these rules are implemented
as follows:

\bldl
r_1: \pfact{ succ(root,root).} { }
r_2: \prule{ succ(X,Y)       } { succ(\_,X),\ d(Y),\ chosen(X, Y).}
r_3:\prule{ chosen(X,Y)     } { succ(\_,X),\ g(X,Y,C),\ \neg diffchoice(X,Y). }
r_4:\prule{ diffchoice(X,Y) } { chosen(X,Y'),\ Y' \neq Y. }
r_5:\prule{ diffchoice(X,Y) } { chosen(X',Y),\ X' \neq X. }
\eldl
\noindent
(Strictly speaking, the {\em chosen} and
{\em diffchoice} predicates should have
been added the subscript $\tt s_2$ for unique
identification. But we dispensed with that, since
there is only one choice rule in the source program
and no ambiguity can occur.)
The diffchoice rules are used to enforce the functional dependencies
$X \rightarrow Y$ and $Y\rightarrow X$ on the chosen tuples.
These conditions can be enforced directly from the stored
table $\tt chosen(X,Y)$ by enforcing the following constraints \footnote{A constraint
is a rule with empty head which is satisfied only if its body is false.}:

\bldl
\prule{ } { chosen(X,Y), \ chosen(X,Y'),\ Y' \neq Y. }
\prule{ } { chosen(X,Y), \ chosen(X',Y),\ X' \neq X. }
\eldl

\noindent that are equivalent to the two rules defining the predicate
$\tt \neg diffchoice$.
Thus, rules $r_4$ and $r_5$ are never executed directly, nor is any
$\tt diffchoice$ atom ever
generated or stored. Thus we can simply  eliminate the diffchoice rules
in the computation of our program $foe(P) = P_C \cup P_D$. In addition, as
previously observed, we can eliminate
the goal  $\tt \neg diffchoice$ from the chosen rules without changing
the definition of $LICO$ (the Lazy Immediate-Consequence Operator
introduced in Definition \ref{LICO-def}).
Therefore, let
$P'_D$ denote $P_D$ after the elimination
of the diffchoice rules, and let $P'_C$
denoted the rules in $P_C$ after the elimination
of their negated diffchoice goals; then, we can express
our LICO computation as follows:

\[
\begin{array}{lll}
\Theta_I    & = & \{ \delta \in  T_{P'_{C}}(I) \setminus  I  \ \mid \
                  I \cup \{ \delta  \} \models FD_P \} \\

\Gamma_P^{lazy}(I) & = & \{ I \cup \{ \delta \} \
                  \mid \ \delta \in \Theta_I \}
                  \ \cup \ \{ I \ \mid \  \Theta_I = \emptyset \}\\

\Psi_P^{lazy}(I)  & =  & \{ \  T_{P'_D}^{\uparrow \omega}(J) \mid\ J \in \Gamma_
{P}(I)\ \}

\end{array}
\]

Various simplifications can be made to this formula.
For program of Example \ref{exp11},
$P'_D$ consists of the exit rule $r_1$, which
only needs to fired once, and of the rule $r_2$,
where the variables in
choice goals are the same as those contained in the head.
In this situation, the head predicate and the $\tt chosen$ predicate can be
stored in the same table and $T_{P'_D}$ is
implemented at no additional cost as part
of the computation of $\tt chosen$.

Consider now the implementation of a table  $\tt chosen_r$.
The  keys for this table
are the left sides of the
choice goals: $\tt X$ and $\tt Y$ for the example at hand.
The data structures needed to support search and insertion on
keys are well-known.
For main memory, we can use hash tables, where searching for a key
value, and  inserting or deleting an entry can be
considered constant-time operations.
Chosen tuples are stored into a table which can be accessed
by means of a set of hash indexes. More specifically, for each
functional dependency $X \rightarrow Y$ there is an hash index
on the attributed specified by the variables in $X$.

\subsection{Naive and Seminaive Implementations}

For Example \ref{exp11}, the application of the LICO
to the empty set,
yields $\Theta_{I_0} = \emptyset$; then,
from the evaluation
of the standard rules we get the set
$\Psi_P^{lazy} (\emptyset) = \tt \{ p(nil, a) \}$.
At the next iteration, we compute $\Theta_{I_1}$ and obtain
all arcs leaving from node $\tt a$.
One of these arcs is chosen and the others are
discarded, as it should be  since they would otherwise
violate the constraint $X \leftrightarrow Y$. This naive
implementation of $\Psi$ generates no redundant computation for
Example \ref{exp11}; similar considerations also hold for the simple
path program of Example \ref{hamiltonian}. In many situations however,
tuples of $\Theta_I$ computed in one iteration, also belong to
$\Theta_I$ in the next iteration, and memorization is less
expensive than recomputation.
Symbolic differentiation techniques similar to
those used in the seminaive fixpoint computation,
can be used to implement this improvement \cite{book97}, as
described below.

We consider the general case, where
a program can have more than one mutually
recursive choice rule and we need to use separate $\tt chosen_r$
tables for each such rule.
For each choice rule $r$, we also store
a table $\tz_r$ with the same attributes as
${\tt chosen}_r$. In $\tz_r$,
we keep the tuples which are
future candidates for the table ${\tt chosen}_r$.

We  update incrementally the content of the tables $\tz_r$ as they were
concrete views, using differential techniques.
In fact, $\Theta_r = \theta_r \sqcup {\tt theta_r}$, where $\tt theta_r$
is the table accumulation for the `old' $\Theta_r$ tuples and
$\theta_r$ is the set of `new' $\Theta_r$ tuples generated
using the differential fixpoint techniques.
Finally, $\Theta_I$ in the LICO is basically
the union of the $\Theta_r$ for the various choice rules $r$.

With ${P'_C}$ be the set of chosen rules in $foe(P)$,
with the $\tt \neg diffchoice$ goal removed;
let $T_r$
denote the immeditate consequence operator for
a rule $r \in P'_C$.
Also, $P'_D$ will
denote  $foe(P)$ after the removal of the $\tt chosen$ rules
and of the $\tt diffchoice$ rules: thus $P'_D$ is $P_D$
whithout the $\tt diffchoice$ rules.

The computation of
$\Psi_P^{lazy}$ can then
be expressed by means of the algorithm reported in Fig.
\ref{algorithm1-figure}.

\begin{figure}\label{algorithm1-figure}
\hrule \vspace*{2mm}
\begin{algorithm}\label{algorithm1} {Semi-naive computation
of a choice model.}\\
\rm
{\bf Input:} Choice program $P$. \\
{\bf Output:} Choice model $I$ for $foe(P)$. \\
{\bf begin}

\begin{enumerate}
\item [0]
Initialization. \\
For every $r \in P'_C$ set  $\tt chosen_r = \tz_r = \emptyset$;\\
Set: $I := T_{P'_{D}}^{\uparrow \omega} (\emptyset)$;
\item [1]
{\bf Repeat}
\begin{enumerate}
\item [(i)]
Select an unmarked arbitrary $r \in P'_C$ and mark $r$
\item [(ii)]
Compute:
$\theta_r = (T_r(I) \setminus \tz_r) \setminus
{\tt conflict}(T_r(I)\setminus \tz_r, {\tt chosen_r})$;
\item [(iii)]
Add $\theta_r$ to $\tz_r$
\end{enumerate}
{\bf Until} $\tz_r \neq \emptyset$ or all rules in $P'_C$ are marked;
\item [2]
{\bf If} $\tz_r = \emptyset$\ {\bf \ Return} $I$;
\item [3]
\begin{enumerate}
\item [(i)]
Select  an arbitrary $x \in \tz_r$, and
move $x$ from $\tz_r$ to ${\tt chosen}_r$;
\item [(ii)]
With $\delta= \{ x \}$, delete from the selected table $\tz_r$
every tuple in ${\tt conflict}( \tz_r, \delta ) $.
\end{enumerate}
\item [4]
Set: $I = T_{P'_{D}}^{\uparrow \omega} (I \cup \delta)$,
then unmark all $P'_C$ rules and resume from Step 1.
\end{enumerate}
{\bf end.}
\end{algorithm}
\vspace*{3mm}
\caption{Semi-naive computation}
\vspace*{2mm}
\hrule
\end{figure}

In fact, the basic computation performed
by our algorithm is operational
translation of $\Psi_P^{lazy}$, enhanced with the
differential computation of $\Theta_r$.
At Step 0, the non-choice rules are computed
strating from the empty set. This corresponds to
the computation of  the non-recursive
rules (exit rules) in all our examples,
but  Examples \ref{matching-ex1}
and \ref{hamiltonian} for which the exit rules are
choice rules and are first computed at Step 1.

In Step 1 and Step 3, of this algorithm, we used the function
 $\tt conflict_r(S,R)$ defined next.
Let $\tt S$ and $\tt R$ be two union-compatible relations, whose
attribute sets contain
the left sides of the choice goals in $r$, i.e., the unique
keys of $\tt chosen_r$
($\tt X$ and $\tt Y$ for the example at hand).
Then, $\tt conflict_r(S,R)$ is the set of tuples in $\tt S$
whose $\tt chosen_r$-key values are also contained in $\tt R$.

Now, Step 1 brings
up to date the content of the $\tz_r$ table,
while ensuring that this does
not contain any tuple
conflicting with tuples in $\tt chosen_r$.

Symbolic differentiation techniques are used to improve
the computation of $T_r(I) \setminus \tz_r$ in recursive rules
at Step 1 (ii) \cite{book97}.    This technique is particularly
simple to apply to a recursive
linear rule where the symbolic differentiation
yields the same rule using, instead of the tuples of the
whole predicate, the delta-tuples computed in the last step.
All our examples but
Examples \ref{hamiltonian} and \ref{tsp} involve linear rule.
The quadratic choice
rule in Example \ref{hamiltonian} is differentiated into
a pair of rules. In all examples, the
delta-tuples are as follows:

(i) The  tuples produced by the exit rules  at Step 0

(ii) The new tuples produced at Step 4 of the last iteration.
For all our examples, Step 4 is a trivial step where
$T_{P'_{D}}^{\uparrow \omega} (I \cup \delta) = I \cup \delta$;
thus $\delta$ is the new value produced at Step 4.

Moreover, if $r$ corresponds to a non-recursive,
as the first rule in Examples \ref{matching-ex1} and \ref{hamiltonian},
then this is only executed once
with  $\tz_r = \emptyset$.

At Step 2, we check the termination condition,
$\Theta_I = \emptyset$, i.e.,
 $\Theta_r = \emptyset$ for all $r$ in $P_C$.

Step 3 (ii) eliminates
from $\tz_r$ all tuples that
conflict with the tuple $\delta$ (including the tuple itself).

Therefore, Algorithm \ref{algorithm1} computes
a stable model for $foe(P)$ since it implements
a differential version of
of the operator $\Psi_P^{lazy}$, and applies this operator until
saturation.

We can now compute the complexity of the example programs
of Section 4. For graphs, we denote by
$n$ and $e$, respectively, the number of their nodes and edges.

\sssection{Complexity of rooted spanning-tree algorithm:
Example \ref{span1}}

The number of chosen tuples is bounded by $O(n)$, while
the number of tuples computed by the evaluation of
body rules is bounded by $O(e)$, since
all arcs connected to the source node
are visited exactly once. Because the cost of generating each
such arc, and
the cost of checking if this is in conflict with a
chosen tuple are $O(1)$, the total cost is $O(e)$.

\sssection{Complexity of single-source reachability algorithm:
Example \ref{reachability}}

This case is very similar to the previous one. The size of
$\tt reach$ is bounded by $O(n)$, and so is the size of the
$\tt chosen$ and $\tz$ relations.
However, in the process of generating $\tt reach$,
all the edges reachable from the source node
$\tt a$ are explored by the algorithm exactly once.
Thus the worst case complexity is $O(e)$.

\sssection{Complexity of simple path: Example \ref{hamiltonian}}

Again, all the edges in the graph will be visited in the worst case,
yielding complexity $O(e)$, where $e= n^2$, according to our
assumption that the graph is complete.
Every arc is visited once and, therefore, the global
complexity is $O(e)$, with $e =n^2$.

\sssection{Complexity of a bipartite matching: Example \ref{matching-ex1}}

Initially all body tuples are inserted into the $\tz$
relation at cost $O(e)$.
The computation terminates in $O(min(n_1,n_2)) = O(n)$ steps, where $n_1$
and $n_2$ are, respectively, the number of
nodes in the left and right parts of the graph.
At each step, one tuple $t$ is selected at cost
$O(1)$ and the tuples conflicting with the selected tuple are deleted.
The global cost of deleting conflicting tuples is $O(e)$, since we
assume that each tuple is accessed in constant time.
Therefore the global cost is $O(e)$.

\sssection{Complexity of linear sequencing of the elements of a set: Example \ref{linear1}}

If $n$ is the cardinality of the domain $\tt d$,
the computation terminates in $O(n)$ steps.
At each step, $n$ tuples are computed, one tuple is
chosen and the remaining tuples are discarded.
Therefore the complexity is $O(n^2)$.


\subsection{Implementation of Choice-least/most  Programs}

In the presence of choice-least  (or choice-most) goals,
the best alternative must be computed, rather
than an arbitrary one chosen at random.
Thus, we introduce a new algorithm, reported in Fig.
\ref{algorithm2-figure},
derived from the algorithm of Fig. \ref{algorithm1-figure}.

Let us consider
the general case where programs could contain
three different kinds of choice rules: (i) choice-least
rules that have one choice least goal, and zero
or more choice goals, (ii)  choice-most rules that have
one choice-most goal and zero or more choice goals, and (iii)
pure choice rules that have one or more choice goals and no
choice-least or choice-most goals.
Then Step 3 (i) in Algorithm \ref{algorithm1} should be modified as
follows:

\begin{quote}
\begin{itemize}
\item [3:]
(i)
If $r$ is a choice-least (choice-most) rule then select a single
 tuple $x \in \tz_r$ with least (most) cost; otherwise ($r$ is
pure choice rule, so) take an arbitrary $x \in \tz_r$.
Move $x$ from $\tz_r$ to ${\tt chosen}_r$;
\end{itemize}
\end{quote}

An additional optimization is however possible,
as discussed next.
Consider for instance Prim's algorithm in Example \ref{prim1}:

Say that $\tz$ contains two tuples  $t_1 = \tt (x, y_1, c_1)$ and
$t_2= \tt (x, y_2, c_2)$. Then, the following properties hold for
Algorithms 1 with Step 3 (i) modified as shown above:
\begin{itemize}
\item
If $\tt c_1 < c_2$ then $t_2$ is not a least-cost tuple,
\item
$t_1$ belongs to $\tt conflict(\tz, \delta)$,
if and only if $t_2$ does.
\end{itemize}
Therefore, the presence of $t_2$ is immaterial to
the result of the computation, and we can
modify our algorithm to ensure that only $t_1$ is
kept in table $\tz$.
This improvement can be implemented by ensuring that
the attribute $\tt Y$ is unique key  for the table $\tz$.
When a new tuple $t'$ is generated  and a tuple with
the same key value is found in $\tz$, the tuple with
the smaller cost value is entered in the table and
the other is discarded. This reduces, the maximum
cardinality of $\tz$ for Prim's and
Dijstra's algorithm from $e$ (number of arcs) to $n$,
(number of nodes).

However, the above considerations are
not valid for rules containing more
than one choice atoms.
For instance, in the greedy TSP program, or the optimal matching
program (Examples \ref{tsp} and \ref{minimum-matching-ex}, respectively),
the  choice rules have
the following choice goals:
\[
\tt  choice((X),(Y)), \  choice((Y),(X)), \ choice\-least((Y),(C))
\]

Say that $\tz$ contains the following tuples:
$t_1 = \tt (x_1, y_1, c_1)$, $t_2= \tt (x_1, y_2, c_2)$
$t_3 = \tt (x_2,y_2,c_3)$, with $c_1 < c_2 <c_3$.
Although, $c_3$ conflicts with $c_2$ and has larger cost
value, it cannot be eliminated, since
it has a chance to be selected later. For instance, if $t_1$ is
selected first then $t_2$ will be eliminated, since it
conflicts with $t_1$. But $t_3$ does not conflict with
$t_1$, and remains, to be selected next.

Thus, the general rule is as follows:

\begin{enumerate}
\item
the union of the left sides of all choice goals is a
unique key for $\tz$,

\item
when a new tuple is inserted and there is a
conflict on the unique key value, retain in $\tz$
only the tuple with the lesser cost.
\end{enumerate}

The above optimization can be carried out by modifying
Step 1 (iii) in Algorithm \ref{algorithm1} as follows:

\begin{quote}
\begin{itemize}
\item [1:]
(iii)
Add each tuple of $\theta_r$ to $\tz_r$; when key conflicts occur,
and $r$ is a choice-least (choice-most) table,
retain the lesser (larger) of the tuples.
\end{itemize}
\end{quote}

Moreover, insertion and deletion of an element from a $\tz$ table can be done
in constant time, since we are assuming that hash indexes are available,
whereas the selection of a least/most cost element is done in linear time.
The selection of the least/most cost element can be done in constant
time by organizing $\tz$ as priority queues.
However the cost
of insertion, deletion of least/most cost element
from a priority queue are now logarithmic, rather
than constant time. Therefore, when using priority queues
we can improve the
performance of our  algorithm
by delaying the
merging $\theta_r$ into $\tt theta_r$ (Step 2), as to allow
the elimination from $\theta_r$ of
tuples conflicting with the
new $\delta$ selected at Step 3.
Therefore, we will
move one tuple from $\theta_r$ to
$\tz_r$ (if $\theta_r \neq \emptyset$),
in Step 1 (iii), and the remaining tuples
at the end of Step 3
(those conflicting with $\delta$ excluded).

\begin{figure}[h]\label{algorithm2-figure}
\hrule \vspace*{2mm}
\begin{algorithm}\label{algorithm2} {Greedy Semi-naive computation of a choice model.}\\
\rm
{\bf Input:} Choice program $P$. \\
{\bf Output:} $I$, a greedy choice model for $P$.\\
{\bf begin}
\begin{enumerate}
\item [0:] Initialization.\\
For every $r \in P'_C$ set  $\tt chosen_r = \tz_r = \emptyset$;\\
Set: $I := T_{P'_{D}}^{\uparrow \omega} (\emptyset)$;
\item [1:]
{\bf Repeat}
\begin{enumerate}
\item [(i)]
Select an unmarked arbitrary $r \in P'_C$ and mark $r$
\item [(ii)]
Compute:
$\theta_r = (T_r(I) \setminus \tz_r) \setminus
{\tt conflict}(T_r(I)\setminus \tz_r, {\tt chosen_r})$;
\item [(iii)]
If $r$ is a choice-least (choice-most) rule then select a single
tuple $x \in \theta_r$ with least (most) cost; otherwise ($r$ is
pure choice rule, so) take an arbitrary $x \in \theta_r$.
Move $x$ from $\theta_r$ to $\tz_r$;
\end{enumerate}
{\bf Until} $\tz_r \neq \emptyset$ or all rules in $P'_{C}$ are marked;
\item [2:]
{\bf If} $\tz_r = \emptyset$\ {\bf \ Return} $I$;
\item [3:]
\begin{enumerate}
\item [(i)]
If $r$ is a choice-least (choice-most) rule then select a single
tuple $x \in \tz_r$ with least (most) cost; otherwise ($r$ is
pure choice rule and) take an arbitrary $x \in \tz_r$.
Move $x$ from $\tz_r$ to ${\tt chosen}_r$;
\item [(ii)]
With $\delta= \{ x \}$, delete from the selected table $\tz_r$
every tuple in ${\tt conflict}( \tz_r, \delta ) $.
\item [(iii)]
Add each tuple of $\theta_r$ to $\tz_r$; when key conflicts occur,
and $r$ is a choice-least (choice-most) table,
retain the lesser (larger) of the tuples.
\end{enumerate}
\item [4:]
Set: $I = T_{P'_{D}}^{\uparrow \omega} (I \cup \delta)$, using the differential
fixpoint improvement,
then unmark all $P'_{C}$ rules and resume from Step 1.
\end{enumerate}
{\bf end.}
\end{algorithm}
\vspace*{-3mm}
\caption{Greedy Semi-naive computation}
\vspace*{2mm}
\hrule
\end{figure}

Observe that the improvement performed in Step 3 reduces the max cardinality
of tables $\tz_r$.
In our Prim's algorithm the size
of the table $\tz$ is reduced from the number of edges $e$ to the number of nodes $n$.
This has a direct bearing on the
performance of our algorithm since the selection
of a least cost tuple is performed $n$ times.
Now, if a linear search is used to find the
least-cost element the global complexity is $O(n \times n)$.
Similar considerations and complexity measures hold for
Dijsktra's algorithm.

In some cases however, the unique key improvement just describe might be
of little or no benefit.
For the TSP program and the optimal matching
program, where the combination of
both end-points is the key for $\tz$,  no benefit
is to be gained since there is at most one edge
between the two nodes. (In a database environment
this might follow from the declaration of unique
keys in the schema, and can thus be automatically
detected by a compiler).

Next, we compute the complexity of the various algorithms, assuming
that the  $\tz$ tables are supported by
simple hash-based indexes,  but there is
no priority queue. The complexities obtained with priority
queues are discussed in the next section.

\sssection{Complexity of Prim's Algorithm: Example \ref{prim1}}

The computation terminates in $O(n)$ steps.
At each step, $O(n)$ tuples are inserted into
the table $\tz$, one least-cost tuple is moved to the
table $\tt chosen$ and conflicting tuples are deleted
from $\tz$.
Insertion and deletion of a tuple is done in constant time,
whereas selection of the least cost tuple is done in linear
time.
Since the size of $\tz$ is bounded by $O(n)$,
the global complexity is $O(n^2)$.

\sssection{Complexity of Dijkstra Algorithm: Example \ref{dijkstra}}

The overall cost is   $O(n^2)$
as for Prim's algorithm.

\sssection{Complexity of sorting the elements of a relation:
Example \ref{sorting} }

At each step, $n$ candidates tuples are generated, one is chosen,
and all tuples are eliminated from $\tz$.
Here, each new $\tt X$ from
$\tt succ$ is matched with every $\tt Y$, even when
differential techniques are used.
Therefore, the cost
is $O(n^2)$.

\sssection{Greedy TSP: Example \ref{tsp}}

The number of steps is equal to $n$. At each step, $n$
tuples are computed by the evaluation of the body of the chosen
rule and stored into the temporary relation $\theta$.
Then,  one tuple with least cost is
selected (Step 1)  and entered in $\tz$
all the remaining tuples are deleted from the relation (Step 3).
The cost of inserting one tuple into the temporary relation is $O(1)$.
Therefore, the global cost is $O(n^2)$.

\sssection{Optimal Matching in a directed graph: Example \ref{minimum-matching-ex}}

Initially all body tuples are inserted into the $\tz$
relation at cost $O(e)$.
The computation terminates in $O(n)$ steps.
At each step, one tuple $t$ with least cost is selected at cost
$O(e)$ and the tuples conflicting with the selected tuple are deleted.
The global cost of deleting conflicting tuples is $O(e)$
(they are accessed in constant time).
Therefore the global cost is $O(e \times n)$.


\subsection{Priority Queues}

In many of the previous algorithms, the dominant cost
is finding the least value in the table $\tz_r$, where
$\tt r$ is a least-choice or most-choice rule.
Priority queues can be used to reduce the overall cost.

A priority queue is a
partially ordered tables where the cost of the $i^{\rm th}$ element
is greater or equal than the cost of the $(i\ div\ 2)^{\rm th}$
element~\cite{AhoHop*74}.
Therefore, our table $\tt theta_r$
can be implemented as a list where each node having position $i$
in the list also contains
(1) a pointer to the next element,
(2) a pointer to the element with position $2 \times i$, and
(3) a pointer to the element with position $i\ div \ 2$.
The cost of finding the least value is constant-time in a
priority queue, the cost of adding or deleting an element
is $log(m)$ where $m$ is the number of the entries in the
queue.

Also, in the implementation of Step 2 (ii), a linear search
can be avoided by adding {\em one search index for
each left side of a choice or choice-least goal}. For instance,
for Dijkstra's algorithm there should be
a search index on $X$, for Prim's on $Y$.
The operation of finding the least cost element in $\theta_r$
can be done during the generation  of the tuples
at no additional cost.
Then we obtain the following complexities:

\sssection{Complexity of Prim's Algorithm: Example \ref{prim1}}

The computation terminates in $O(n)$ steps and the size
of the priority queue is bounded by $O(n)$.
The number of candidate tuples is bounded by $O(e)$.
Therefore, the global cost is bounded by $O(e \times log\, n)$.

\sssection{Complexity of Dijkstra Algorithm: Example \ref{dijkstra}}

The overall cost is   $O(e \times log\, n)$
as for Prim's algorithm.

\sssection{Complexity of sorting the elements of a relation:
Example \ref{sorting} }

The number of steps is equal to $n$. At each step, $n$
tuples are computed, one is stored into $\tz$ and next
moved to $\tt chosen$ while
all remaining tuples are deleted from $\theta$.
The cost of each step is $O(n)$ since deletion of a
tuple from $\theta$ is constant time.
Therefore, the global cost
is $O(n^2)$.

\sssection{Greedy TSP: Example \ref{tsp}}

Observe that
$\tz_r$ here contains at most one tuple. The addition
of the first tuple into an empty priority queue, $\tz_r$,
and the deletion of the last tuple from it are constant
time operations. Thus  the overall cost is the same as
that without a priority queue: i.e. the global cost is $O(n^2)$.

\sssection{Optimal Matching in a bipartite graph:
Example \ref{minimum-matching-ex}}

Initially, all body tuples are inserted into the $\tz$
relation at cost $O(e \times log\, e)$.
The computation terminates in $O(n)$ steps.
At each step, one tuple $t$ is selected and all remaining
tuples conflicting with $t$ are deleted (the conflicting tuples here
are those arcs having the same node as source or end node of the arc).
The global number of extractions from the priority queue is
$O(e)$.
Therefore, the global complexity is $O(e \times log\, e)$.

\vspace{3mm}
Observe that, using a priority queues, an asymptotically optimum
performance~\cite{AhoHop*74} has been achieved for all problems,
but that of sorting the elements of a domain,
Example~\ref{sorting}. This problem is
considered in the next section.


\subsection{Discussion}

A look at the
structure of the  program in Example~\ref{sorting}
reveals that at the beginning of each step
a new set of $(x,y)$ pairs is generated for $\tz$
by the two goals $\tt succ(\_, X), d(Y)$ which define a
Cartesian product. Thus, the computation can be represented
as follows:
\[
\Theta = \pi_2 {\tt succ} \times {\tt d}
\]
where $\tt succ$ and $\tt d$ are the relations containing their
homonymous predicates. We can also represent $\theta$
and $\tz$ as  Cartesian products:
\[
\theta = (\pi_2 \delta \times {\tt d}) \setminus {\tt chosen} =
\pi_2 \delta \times (d \setminus \pi_2 {\tt chosen})
\]

Therefore, the key to obtaining
an efficient implementation here consists in storing
only the second column of the $\tz$ relation, i.e.:
\[
\pi_2 \tz = {\tt d} \setminus \pi_2 {\tt chosen}
\]

The operation of selecting a least-cost tuple from $\tz$
now reduces to that of selecting a least-cost tuple from
$\pi_2 \tz$, which  therefore should be  implemented as
a priority queue.

Assuming these modications, we can now recompute the
complexity of our Example 3, by observing that
all the elements in $\tt d$ are added
to $\pi_2 \tz$ once at the first iteration.
Then each successive iteration eliminates one element
from this set.
Thus, the overall complexity
is linear in the number of nodes.
For Example 11, the complexity is $O(n \times log\, n)$ if
we assume that a priority queue is kept for $\pi_2 \tz$
Thus we obtaine the optimal complexities.

No similar improvement is applicable to
the other examples,
where the rules do not compute the Cartesian product of two relations.
Thus, this additional improvement could also be incorporated
into a smart compiler, since it is possible to
detect from the rules whether $\tz$ is in fact the Cartesian
product of its two subprojections. However this
is not the only alternative  since many existing
deductive database systems  provide the user with
enough control to implement this, and other
differential  improvements previously discussed, by coding them
into the program. For instance, the {\LDL}++ users could
use XY-stratified programs for this
purpose \cite{ZAO93};  similar programs can be used in other
systems \cite{aditi}.


\section{Conclusion}
This paper has introduced a logic-based approach for
the design and implementation of greedy algorithms.
In a nutshell, our design approach is as follows: (i)
formulate the all-answer solution for the problem at
hand (e.g., find all the costs of all paths from
a source node to other nodes), (ii)
use choice-induced FD constraints to
restrict the original logic program to the non-deterministic
generation of a single answers (e.g., find a cost from the
source node to each other node), and (iii) specialize
the choice goals with preference annotations to force
a greedy heuristics upon the generation of single
answers in the choice-fixpoint algorithm (thus computing
the least-cost paths). This approach yields conceptual
simplicity and simple programs; in fact it has been
observed that our programs are often similar to
pseudo code expressing the same problem in a procedural
language. But our approach offers additional advantages,
including  a formal logic-based semantics and a clear
design method, implementable by a compiler,
to achieve optimal implementations for our
greedy programs. This method is based on
\begin{itemize}
\item
The use of $\tt chosen$ tables and $\tt theta$ tables, and of
differential techniques to support the second kind of table as
a concrete view. The actual structure of $\tt theta$ tables, their search
keys and unique keys are determined by the choice and choice-least
goals, and the join dependencies implied by the structure
of the original rule.
\item
The use of priority queues for expediting the finding of
extrema values.
\end{itemize}
Once these general guidelines are followed (by a user
or a compiler) we obtain an implementation that achieves
the same asymptotic complexity
as procedural languages.

This paper provides a refined example of the power of
Kowalski's seminal idea: {\em algorithms = logic + control}.
Indeed, the logic-based approach here proposed covers all aspects
of greedy algorithms, including (i) their
initial derivation using rules with choice goals, (ii) their
final formulation by choice-least/most goals,
(iii) their  declarative stable-model semantics, (iv) their
operational (fixpoint)  semantics, and   finally (v)
their optimal implementation  by syntactically derived
data structures and indexing methods.
This vertically integrated, logic-based,
analysis and design methodology represents a significant
step forward with respect to previous logic-based approaches
to greedy algorithms
(including those we have
proposed in the past \cite{GreZan*92,GanGre*95}).

\paragraph{Acknowledgement}
The authors would like to thank D. Sacc\`a for
many hepfull discussions and suggestions. The referees deserve
credit for many improvements.



\label{lastpage}


\begin{thebibliography}{}


\bibitem[\protect\citename{Abiteboul \emph{et al.}, }1994]{AHV94}
Abiteboul, S., Hull, R. and Vianu, V. (1994)
\emph{Foundations of Databases}.
Addison-Wesley.

\bibitem[\protect\citename{Abiteboul and Vianu, }1991]{AbiVia91}
Abiteboul, S. and Vianu, V. (1991)
Datalog Extensions for Databases Queries and Updates.
\emph{Journal of Computer and System Science}, 43~(1): pp. 62--124.
Academic Press.

\bibitem[\protect\citename{Aho \emph{et al.}, }1974]{AhoHop*74}
Aho, A.V., Hopcropt, J.E. and Ullman, J.D. (1974)
\emph{The Design and Analysis of Computer Algorithms}.
Addison-Wesley.

\bibitem[\protect\citename{Apt \emph{et al.}, }1988]{AptBla*88}
Apt, K.R., Blair, H.A. and Walker, A. (1988)
Towards a theory of declarative knowledge,
In J. Minker (editor)
\emph{Foundations of Deductive Databases and Logic Programming}, pp. 89--148.
Morgan Kaufmann.

\bibitem[\protect\citename{Chandra and Harel, }1982]{ChH82}
Chandra, A. and Harel, D. (1982)
Structure and Complexity of Relational Queries,
\emph{Journal of Computer and System Science},  25~(1): 99--128.
Academic Press.


\bibitem[\protect\citename{Dietrich, }1992]{Die92}
Dietrich, S. W. (1992)
Shortest Path by Approximation in Logic Programs,
\emph{ACM Letters on Programming Languages and Systems}, 1~(2): pp. 119--137.
ACM Press.

\bibitem[\protect\citename{Ganguly \emph{et al.}, }1995]{GanGre*95}
Ganguly, S., Greco, S. and Zaniolo, C. (1995)
Extrema Predicates in Deductive Databases.
\emph{Journal of Computer and System Science}, 51~(2): pp. 244--259.
Academic Press.

\bibitem[\protect\citename{Gelfond and Lifschitz, }1988]{gelf-lifs-88}
Gelfond, M. and Lifschitz, V. (1988)
The stable model semantics of logic programming.
In \emph{Proceedings Fifth Internernational Conference on Logic Programming},
Jerusalem, Israel, June 18-20, pp. 1070--1080.
MIT Press.

\bibitem[\protect\citename{Giannotti \emph{et al.}, }1991]{GiaPed*91}
Giannotti, F., Pedreschi, D., Sacc\`{a}, D. and Zaniolo, C. (1991)
Nondeterminism in deductive databases.
In \emph{Proceedings of the 2nd International Conference on Deductive
and Object-Oriented Databases},
Munich, Germany, December 16-18.
Lecture Notes in Computer Science, Vol. 566, pp. 129--146.
Springer.

\bibitem[\protect\citename{Giannotti \emph{et al.}, }1999]{GPZ96}
Giannotti, F., Pedreschi, D. and Zaniolo, C. (2000) Semantics and
Expressive Power of Non-Deterministic Constructs in Deductive
Databases, \emph{Journal of Computer and System Science} (to appear).
Academic Press.

\bibitem[\protect\citename{Greco and Zaniolo, }1998]{report}
Greco, S. (1999)
Dynamic Programming in Datalog with Aggregates.
\emph{IEEE Transaction on Knowledge Engineering}, 11(2): pp. 265--283.
IEEE Computer Society.

\bibitem[\protect\citename{Greco \emph{et al.}, }1995]{GreSac*95}
Greco S., Sacc\`{a}, D. and Zaniolo, C. (1995)
DATALOG Queries with Stratified Negation and Choice: from $P$ to $D^P$.
In \emph{Proceedings Fifth International Conference on Database Theory}
Lecture Notes in Computer Science, Vol. 893, pp. 81--96.
Springer.

\bibitem[\protect\citename{Greco \emph{et al.}, }1992]{GreZan*92}
Greco, S., Zaniolo, C. and Ganguly, S. (1992)
Greedy by Choice.
In \emph{Proceedings of the 11th ACM Symposium on Principles of  Database Systems},
June 2-4, 1992, San Diego, California, pp. 105--113.
ACM Press.

\bibitem[\protect\citename{Greco and Zaniolo, }1997]{GreZan97}
Greco, S. and Zaniolo, C. (1998)
Greedy Algorithms in Datalog with Choice and Negation.
In \emph{Proceedings International Joint Conference and Symposium on Logic Programming},
Manchester, UK, 15-19 June 1998, pp. 294--309.
MIT Press.

\bibitem[\protect\citename{Krishnamurthy and Naqvi, }1988]{KriNaq88}
Krishnamurthy, R. and Naqvi, S. (1988)
Non-deterministic choice in {Datalog}.
In \emph{Proceedings of the 3rd International Conference on Data and Knowledge Bases},
June 28-30, 1988, Jerusalem, Israel, pp. 416--424.
Morgan Kaufmann.

\bibitem[\protect\citename{Marek and Truszczynski, }1991]{MarTru91}
Marek, W. and Truszczynski, M. (1991)
Autoepistemic Logic.
\emph{\em Journal of ACM}, 38(3): pp. 588--619.
ACM Press.

\bibitem[\protect\citename{Moret and Shapiro, }1993]{MorSha91}
Moret, B. M. E. and Shapiro, H.D. (1993)
\emph{Algorithms from P to NP}.
Benjamin Cummings.

\bibitem[\protect\citename{Papadimitriou and Steiglitz, }1975]{PapLew75}
Papadimitriou, C. and Steiglitz, K. (1975)
\emph{Combinatorial Optimization: Algorithms and Complexity.}
Englewood Cliff, N.J., Prentice Hall.

\bibitem[\protect\citename{Przymusinski, }1988]{Prz88}
Przymusinski, T. (1988)
On the declarative and procedural semantics of stratified
deductive databases.
In J. Minker (editor)
\emph{Foundations of Deductive Databases and Logic Programming}, pp. 193--216.
Morgan-Kaufman.

\bibitem[\protect\citename{Przymusinska and Przymusinski, }1988]{PrzPrz88}
Przymusinska, A. and Przymusinski, T. (1988)
Weakly Perfect Model Semantics for Logic Programs.
In \emph{Proceedings of the Fifth International Conference and Symposium on Logic Programming},
Seattle, Washington, August 15-19, pp. 1106--1122.
MIT Press.


\bibitem[\protect\citename{Sudarshan and Ramakrishnan, }1991]{RosSag92}
Ross, K. A. and Sagiv, Y. (1992)
Monotonic Aggregation in Deductive Databases.
In \emph{Proceedings of the 11th ACM Symposium on Principles of Database Systems},
June 2-4, 1992, San Diego, California, pp. 114--126.
ACM Press.

\bibitem[\protect\citename{Sacc\`{a} and Zaniolo, }1990]{SacZan90}
Sacc\`{a}, D. and Zaniolo, C. (1990)
Stable models and non-determinism in logic programs with negation.
In \emph{Proceedings of the Ninth ACM Symposium on Principles of Database Systems},
April 2-4, 1990, Nashville, Tennessee, pp. 205--217.
ACM Press.


\bibitem[\protect\citename{Sudarshan and Ramakrishnan, }1991]{SudRam91}
Sudarshan, S. and Ramakrishnan, R. (1991)
Aggregation and relevance in deductive databases.
In \emph{Proceedings of the 17th Conference on Very Large Data Bases},
September 3-6, Barcelona, Catalonia, Spain, pp. 501--511.
Morgan Kaufmann.

\bibitem[\protect\citename{Ullman, }1989]{Ull88}
Ullman, J. D. (1989)
\emph{Principles of Data and Knowledge-Base Systems}, Vol. 1 \& 2.
Computer Science Press.

\bibitem[\protect\citename{Vaghani \emph{et al.}, }1994]{aditi}
Vaghani, J., Ramamohanarao, K., Kemp, D. B., Somogyi, Z., Stuckey,
P. J., Leask, T. S. and Harland, J. (1994)
The Aditi deductive database system.
\emph{The VLDB Journal}, 3(2): pp. 245--288.
Springer.

\bibitem[\protect\citename{Van Gelder \emph{et al.}, }1991]{GelRos*91}
Van~Gelder, A., Ross, K.A. and Schlipf, J.S. (1991)
The well-founded semantics for general logic programs.
\emph{Journal of ACM}, 38(3): pp. 620--650.
ACM Press.

\bibitem[\protect\citename{Van Gelder, }1993]{Gel93}
Van~Gelder, A. (1993)
Foundations of Aggregations in Deductive Databases
In \emph{Proceedings of the 3rd International Conference On Deductive and Object-Oriented Databases},
Phoenix, Arizona, USA, December 6-8.
Lecture Notes in Computer Science, Vol. 760, pp. 13--34.
Springer.

\bibitem[\protect\citename{Zaniolo \emph{et al.}, }1993]{ZAO93}
Zaniolo, C., Arni, N. and Ong, K. (1993)
Negation and Aggregates in Recursive Rules: the {\LDL}++ Approach,
In \emph{Proceedings of the 3rd International Conference on Deductive and Object-Oriented Databases},
Phoenix, Arizona, USA, December 6-8.
Lecture Notes in Computer Science, Vol. 760, pp. 204--221.
Springer.

\bibitem[\protect\citename{Zaniolo \emph{et al.}, }1997]{book97}
Zaniolo C., S. Ceri, C. Faloutsos, V.S.
Subrahmanian and R. Zicari,
\emph{Advanced Database Systems},
Morgan Kaufmann.

\end{thebibliography}
\end{document}